\documentclass[nofootinbib,twocolumn]{revtex4} 

\usepackage[dvips]{graphicx}

\begin{document}


\title{Interacting vacuum energy in the dark sector\footnote{To appear in AIP Conf. Proc. of CosmoSur II - Cosmology and Gravitation in the Southern Cone, Santiago, Chile, May 2013.}}

\author{L. P. Chimento$^1$\footnote{chimento.df.uba.ar} and S. Carneiro$^2$\footnote{saulo.carneiro@pq.cnpq.br}}

\affiliation{$^1$ Departamento de F\'{\i}sica, Facultad de Ciencias
Exactas y Naturales, Universidad de Buenos Aires and IFIBA,
CONICET, Cuidad Universitaria, Buenos Aires 1428,  Argentina\\ $^2$
Instituto de F\'{\i}sica, Universidade Federal da Bahia, 40210-340, Salvador, BA,
Brazil}


\date{1 April 2014}

\begin{abstract}
We analyse three cosmological scenarios with interaction in the dark sector, which are particular cases of a general expression for the energy flux from vacuum to matter. In the first case the interaction leads to a transition from an unstable de Sitter phase to a radiation dominated universe, avoiding in this way the initial singularity. In the second case the interaction gives rise to a slow-roll power-law inflation. Finally, the third scenario is a concordance model for the late-time universe, with the vacuum term decaying into cold dark matter. We identify the physics behind these forms of interaction and show that they can be described as particular types of the modified Chaplygin gas.
\end{abstract}

\maketitle

\section{Introduction}

The interaction between the two components of the dark sector of the Universe is not only allowed by Einstein's equations but also leads to interesting scenarios for both the primordial and late-time phases of the expansion \cite{chimento1,zimdahl,sergio,elcio}. In this letter we review three of these scenarios and show that they are particular cases of a general expression for the energy flux from vacuum to the matter sector. In the first of them, the vacuum-matter interaction leads to a transition from an unstable de Sitter phase to a radiation dominated universe, avoiding in this way the initial singularity. In the second case, the interaction gives rise to an inflationary expansion without inflaton, with desired features of standard inflationary models. Finally, the third scenario is a concordance model for the late-time universe, with the vacuum term decaying into cold dark matter. In the three cases we identify the physics behind these particular forms of interaction. We also show that they can be interpreted as three particular types of the modified Chaplygin gas.

\section{Interacting models}

Let us consider a FLRW space-time with two energy components, namely a vacuum term with equation of state $p_x = (\gamma_x - 1) \rho_x$ and a matter component with equation of state $p_m = (\gamma_m - 1)  \rho_m$, where $\rho_i$ and $p_i$ ($i=x,m$) are the corresponding energy densities and pressures, respectively. The Friedmann equations are given by
\begin{equation} \label{Friedmann}
3H^2 = \rho_m + \rho_x,
\end{equation}
\begin{equation} \label{conservation}
\dot{\rho}_m + \dot{\rho}_x + 3H(\gamma_m \rho_m + \gamma_x \rho_x) = 0,
\end{equation}
where the dot means derivative with respect to the cosmological time $t$, and $H = \dot{a}/a$ is the Hubble parameter. Introducing the total energy density $\rho = \rho_m + \rho_x$, redefining the time variable by $d\tau = 3 H dt = 3da/a$, and choosing $\gamma_x=0$ for the vacuum term, it is easy to obtain
\begin{equation} \label{conservation2}
\rho' + \gamma_m \rho_m = 0,
\end{equation}
where the prime means derivative with respect to $\tau$.
On the other hand, derivating (\ref{Friedmann}) with respect to time we can derive
\begin{equation} \label{H}
2\dot{H} = \rho'.
\end{equation}

The interaction between the two components can be introduced by splitting the conservation equation (\ref{conservation2}) into
\begin{equation} \label{matter}
\rho'_m + \gamma_m \rho_m = - Q,
\end{equation}
\begin{equation} \label{vacuum}
\rho'_x = Q,
\end{equation}
where $Q$ is a generic expression of the energy flux, which is negative if the flux is from vacuum to matter. Multiplying (\ref{matter}) by $\gamma_m$ and using (\ref{conservation2}) we have 
\begin{equation} \label{general}
\rho'' + \gamma_m \rho' = \gamma_m Q.
\end{equation}
Assuming that $Q = Q (\rho,\rho')$, we then have a second-order equation for $\rho$, which determines the dynamics for a given $Q$.

\section{\lowercase{de} Sitter phase transition}

Let us first take
\begin{equation} \label{Q1}
Q = \frac{2}{3} \rho \rho'.
\end{equation}
The source equation (\ref{general}) is integrated to give
\begin{equation} \label{case1}
\rho' + \gamma_m \rho = \frac{\gamma_m}{3} \rho^2 + C_1,
\end{equation}
where $C_1$ is an integration constant which will be taken as zero. If we now write $\rho_x = \rho - \rho_m$ and use (\ref{conservation2}) and (\ref{case1}), we obtain
\begin{equation} \label{H4}
\rho_x = 3H^4.
\end{equation}
On the other hand, from (\ref{H}) and (\ref{case1}) we have, by doing $\gamma_m = 4/3$ (i.e. $p_m = \rho_m/3$), the evolution equation
\begin{equation} \label{evolution}
\dot{H} = -2H^2 + 2H^4.
\end{equation}

\begin{figure}
\includegraphics{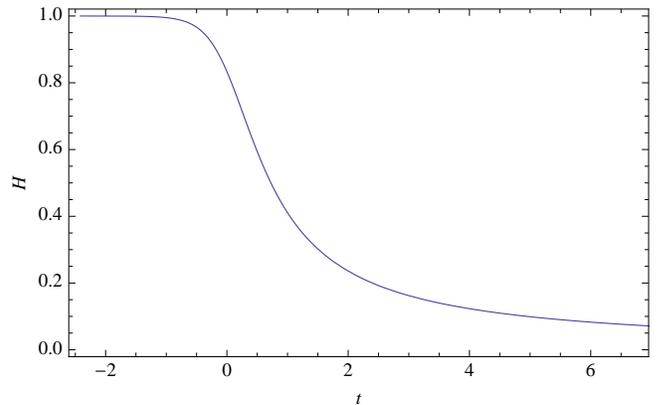}
\caption{\label{Fig1} The Hubble parameter as a function of time for $\Lambda=3H^4$}
\end{figure}

Equations (\ref{H4}) and (\ref{evolution}) are noteworthy. The former is the expected contribution of free conformal fields to vacuum density in de Sitter space-time. It is valid in particular in the high energy limit, when field interactions and masses can be neglected. Since in de Sitter background we have $\rho = \rho_x = 3H^2$, equation (\ref{H4}) leads to $H = 1$, that is, we are dealing with a Planck-scale solution\footnote{If we multiply (\ref{Q1}) by a constant factor, the scale changes by the same factor. Nevertheless, the behavior of our solution is not altered, except for a simple re-scaling of time.}. Equation (\ref{evolution}), on the other hand, has two possible solutions. The first is simply $H = 1$, which is an ethernal de Sitter universe. However, this solution is unstable, which is clear because of the existence of a second solution for (\ref{evolution}), namely \cite{saulo1,tavakol}
\begin{equation} \label{evolution2}
2t = \frac{1}{H} - \tanh^{-1} H,
\end{equation}
where a constant of integration was conveniently chosen. This solution is depicted in Fig. \ref{Fig1}. It describes a phase transition from an initial, semi-ethernal de Sitter phase to a radiation dominated universe, during which the vacuum term decays into relativistic matter. This result can also be seen from the solution of Eq. (\ref{case1}) for $C_1=0$, the branch of which, defined for any value of the energy density, is given by 
\begin{equation} \label{scase1}
\rho=\frac{3}{1+c\,a^{4}},
\end{equation}
where $c$ is a positive integration constant. The particular solution obtained for $c=0$ gives rise to the unstable de Sitter solution $H=1$, which in turns decays into relativistic matter with $\rho\approx 3/ca^{4}$.
When this solution departures significantly from de Sitter, the vacuum density is not expected to be (\ref{H4}) any more, because the backreaction of the particle creation should also be taken into account. This leads us to our second case.

\section{Inflation}

Let us now take
\begin{equation} \label{Q2}
Q = \frac{\gamma}{3\gamma_m} \rho',
\end{equation}
where $\gamma$ is a positive constante. From (\ref{general}) we obtain
\begin{equation} \label{case2}
\rho' + \left( \gamma_m - \frac{\gamma}{3} \right) \rho = C_1,
\end{equation}
where, as before, $C_1$ is an integration constant which can be made zero. Since $\rho_x = \rho - \rho_m$, we now have, by using again (\ref{conservation2}) and $\gamma_m = 4/3$,
\begin{equation} \label{inflation}
\rho_x = \frac{3\gamma}{4} H^2.
\end{equation}
Now, with the help of (\ref{Q2}) and (\ref{conservation2}), equation (\ref{matter}) can be rewritten in the form
\begin{equation} \label{boltzmann4}
\frac{1}{a^4} \frac{d}{dt} \left( \rho_m a^4 \right) = \gamma H \rho_m.
\end{equation}
This equation shows that the vacuum scaling (\ref{inflation}) corresponds to creation of relativistic particles at a rate $\gamma H$.

The solution of (\ref{Friedmann}), (\ref{inflation}) and (\ref{boltzmann4}) is given by \cite{saulo2}
\begin{equation} \label{inflation2}
H = \frac{1}{\epsilon t},
\end{equation}
where $\epsilon = (4 - \gamma)/2$ and an integration constant was conveniently chosen. For $\gamma \approx 4$ (i.e. $\epsilon \ll 1$) we have a slow-roll power-law inflationary solution. Indeed, from (\ref{inflation2}) we can show that $\epsilon = -\dot{H}/H^2$, that is, $\epsilon$ coincides with the first slow-roll parameter. As $\epsilon$ is constant, the second slow-roll parameter is $\delta = -\epsilon$, the scalar spectral index is
$n_s = 1 - 2\epsilon$, and the tensor/scalar ratio is $r = 9 \epsilon$.
Assuming that inflation lasts for about $60$ e-folds, as required to solve the horizon problem, we have $\epsilon \approx 1/60$, leading to $n_s \approx 0.97$ and $r \approx 0.15$.

\section{Late-time limit}

Finally, let us take
\begin{equation} \label{Q3}
Q = \frac{\Gamma}{\gamma_m} \frac{\rho'}{\sqrt{3\rho}},
\end{equation}
where $\Gamma$ is a positive constant. The integration of (\ref{general}) gives
\begin{equation} \label{case3}
\rho' + \gamma_m \rho = \frac{2\Gamma}{\sqrt{3}} \sqrt{\rho} + C_1.
\end{equation}
Writing again $\rho_x = \rho - \rho_m$, using (\ref{conservation2}) and (\ref{case3}), choosing $C_1 = 0$ and taking $\gamma_m = 1$ (i.e. $p_m = 0$), we have
\begin{equation} \label{linear}
\rho_x = 2\Gamma H.
\end{equation}
Again, with the help of (\ref{Q3}) and (\ref{conservation2}), equation (\ref{matter}) can be put in the form
\begin{equation} \label{boltzmann1}
\frac{1}{a^3} \frac{d}{dt} \left( \rho_m a^3 \right) = \Gamma \rho_m.
\end{equation}
We see that, in this case, non-relativistic matter is created with a constant rate $\Gamma$. Dividing (\ref{linear}) by $3H^2$, we have
\begin{equation} \label{Gamma}
\Gamma = \frac{3}{2} \left( 1 - \Omega_m \right) H,
\end{equation}
where $\Omega_m$ is the relative matter density. By taking present values for $H$ and $\Omega_m$ ($\approx 1/3$), we have $\Gamma \approx H_0$.

The system of equations (\ref{Friedmann}), (\ref{linear}) and (\ref{boltzmann1}) has a simple solution given by \cite{borges}
\begin{equation} \label{latetime}
a = C \left[ e^{\Gamma t} - 1 \right]^{2/3},
\end{equation}
where $C$ is an integration constant. It reduces to the Einstein-de Sitter solution for early times, and tends to a de Sitter universe in the asymptotic future. It has been shown in previous works that this solution presents good concordance when tested against current observations \cite{alcaniz}.

\section{Concluding remarks}

The three cases above can be written in a unified form. Indeed, by using the pairs (\ref{Q1})-(\ref{H4}), (\ref{Q2})-(\ref{inflation}) and (\ref{Q3})-(\ref{linear}), we can write the energy flux as
\begin{equation} \label{Q}
Q = - \alpha \,\frac{\rho_x \rho'}{\rho},
\end{equation}
where $\alpha = -2, -1, -1/2$ in the first, second and latter cases, respectively. On the other hand, from (\ref{Friedmann}) and (\ref{conservation2}) it is easy to show that
\begin{equation}
\rho_x = \rho + \frac{\rho'}{\gamma_m}.
\end{equation}
Substituting this result into (\ref{Q}), we obtain
\begin{equation}
Q = - \alpha \left( \rho' + \frac{\rho'^2}{\gamma_m \rho} \right).
\end{equation}
This last equation has the general form of what has been called a homogeneous non-linear interaction in the dark sector \cite{chimento2}. 

It is also noteworthy that in the three cases the dark sector can be described in a unified way as a modified Chaplygin gas. Indeed, equations (\ref{H4}), (\ref{inflation}) and (\ref{linear}) can be put in the general form
\begin{equation}
\rho_x = A \rho^{-\alpha},
\end{equation}
with $\alpha$ defined as above. Now, by using
\begin{eqnarray}
\rho &=& \rho_x + \rho_m,\\ p &=& -\rho_x + (\gamma_m -1) \rho_m,
\end{eqnarray}
we can show that
\begin{equation}
p = (\gamma_m -1) \rho - \frac{B}{\rho^{\alpha}},
\end{equation}
with $B = \gamma_m A$. In the late-time case of non-relativistic dark matter ($\gamma_m = 1$), the dark sector is  described by a generalised Chaplygin gas with negative $\alpha$ \cite{fabris}, also named extended Chaplygin gas \cite{chimento3}.

\end{document}